\documentclass[reprint,floatfix,amsmath,amssymb,aps,pra,superscriptaddress]{revtex4-2}
\usepackage{comment}
\usepackage{tikz}
\usepackage{amsmath}
\usepackage{amssymb}
\usepackage{amsthm}
\usepackage{amsfonts}
\usepackage{graphicx}
\usepackage{bbm}
\usepackage{url}
\usepackage{stmaryrd}
\usepackage{hyperref}
\usepackage{varioref}
\usepackage{cleveref}
\usepackage{braket}
\usepackage{ upgreek }
\usepackage{dsfont}
\usepackage{makecell}
\usepackage{natbib}
\usepackage{fancyref} 
\usepackage{algorithm}
\usepackage{algpseudocode}
\usepackage[utf8]{inputenc}
\usepackage[T1]{fontenc}

\newcommand{\identity}{\ensuremath{\mathds{1}}}

\usepackage{float}
\newfloat{algorithm}{t}{lop}

\newcommand{\abs}[1]{\left| #1 \right|}

\usepackage{hyperref}

\usepackage{tikz}
\usepackage{ifthen}
\usepackage{pgfplots}
\pgfplotsset{compat=1.9}
\usetikzlibrary{shapes,arrows}
\usetikzlibrary{positioning}
\usetikzlibrary{shapes.geometric}

\theoremstyle{plain}
\newtheorem{theorem}{Theorem}

\newtheorem{cor}{Corollary}

\theoremstyle{remark}

\newcommand*{\fancyrefthmlabelprefix}{thm}
\newcommand*{\fancyreflemlabelprefix}{lem}
\newcommand*{\fancyrefcorlabelprefix}{cor}
\newcommand*{\fancyrefdefilabelprefix}{defi}
\frefformat{plain}{\fancyreflemlabelprefix}{lemma\fancyrefdefaultspacing#1}
\Frefformat{plain}{\fancyreflemlabelprefix}{Lemma\fancyrefdefaultspacing#1}
\frefformat{plain}{\fancyrefthmlabelprefix}{theorem\fancyrefdefaultspacing#1}
\Frefformat{plain}{\fancyrefthmlabelprefix}{Theorem\fancyrefdefaultspacing#1}
\frefformat{plain}{\fancyrefcorlabelprefix}{corollary\fancyrefdefaultspacing#1}
\Frefformat{plain}{\fancyrefcorlabelprefix}{Corollary\fancyrefdefaultspacing#1}
\frefformat{plain}{\fancyrefdefilabelprefix}{definition\fancyrefdefaultspacing#1}
\Frefformat{plain}{\fancyrefdefilabelprefix}{Definition\fancyrefdefaultspacing#1}
\newcommand*{\fancyrefalglabelprefix}{alg}
\newcommand*{\frefalgname}{algorithm}
\newcommand*{\Frefalgname}{Algorithm}
\frefformat{plain}{\fancyrefalglabelprefix}{%
  \frefalgname\fancyrefdefaultspacing#1%
}%
\Frefformat{plain}{\fancyrefalglabelprefix}{%
  \Frefalgname\fancyrefdefaultspacing#1%
}%

\newcommand*{\fancyrefapplabelprefix}{app}
\newcommand*{\frefappname}{appendix}
\newcommand*{\Frefappname}{Appendix}
\frefformat{plain}{\fancyrefapplabelprefix}{%
  \frefappname\fancyrefdefaultspacing#1%
}%
\Frefformat{plain}{\fancyrefapplabelprefix}{%
  \Frefappname\fancyrefdefaultspacing#1%
}%

\def\Block[#1,#2,#3]{

\def\r{0.3};

\ifthenelse{\NOT #3=0}{
\fill [#2] (-0.5,-0.5) rectangle ({#1-0.5},0.5);
}

\foreach \n in {1,...,#1}{

\shade[shading=ball, ball color=coolblue] ({\n-1},0) circle (\r);

}
}

\definecolor{Green}{HTML}{00AD69}  

\def\beq{\begin{equation}}
\def\eeq{\end{equation}}
\def\bq{\begin{quote}}
\def\eq{\end{quote}}
\def\ben{\begin{enumerate}}
\def\een{\end{enumerate}}
\def\bit{\begin{itemize}}
\def\eit{\end{itemize}}

\def\l|{\left|}
\def\r|{\right|}

\newcommand{\cD}{\mathcal{D}}

\newcommand{\cL}{\mathcal{L}}

\usepackage{enumitem}

\newcommand{\tr}{{\operatorname{tr}}}

\newcommand{\Id}{\mathds{1}}

\newcommand{\norm}[1]{\left\|#1\right\|}

\newcommand{\cH}{\mathcal{H}}

\newcommand{\eps}{\epsilon}

\definecolor{Green}{HTML}{00AD69}  
\definecolor{coolblue}{RGB}{0,102,102}
\definecolor{coolgreen}{RGB}{0,153,153}
\definecolor{shinyblue}{RGB}{0,255,255}
\definecolor{lightblue}{RGB}{102,210,255}
\definecolor{lightpurple}{RGB}{140,30,255}
\definecolor{lightpink}{RGB}{204,0,204}
\definecolor{midblue}{RGB}{0,102,204}
\definecolor{midpink}{RGB}{153,0,153}
\definecolor{darkblue}{RGB}{0,0,153}
\definecolor{cyan}{RGB}{0,204,204}
\definecolor{lightgreen}{RGB}{0,255,128}
\definecolor{midgreen}{RGB}{0,204,0}
\definecolor{darkgreen}{RGB}{0,102,51}
\definecolor{midyellow}{RGB}{204,204,0}
\definecolor{darkyellow}{RGB}{153,153,0}
\definecolor{darkpurple}{RGB}{102,0,102}
\definecolor{darkred}{RGB}{102,0,0}
\definecolor{neworange}{RGB}{255,153,51}

\begin{document}

\title{Rapid thermalization of spin chain commuting Hamiltonians}

\author{Ivan Bardet}
\email{ivan.bardet@inria.fr}
\affiliation{Inria Paris, 75012 Paris, France}
\author{{\'A}ngela Capel}
\email{angela.capel@uni-tuebingen.de}
\affiliation{Department of Mathematics, Technische Universit\"at M\"unchen, 85748 Garching, Germany}
\affiliation{Munich Center for Quantum Science and Technology (MCQST), 80799 München, Germany}
\affiliation{Fachbereich Mathematik, Universität Tübingen, 72076 Tübingen, Germany}
\author{Li Gao}
\email{lgao20@central.uh.edu}
\affiliation{Department of Mathematics, Technische Universit\"at M\"unchen, 85748 Garching, Germany}
\affiliation{Munich Center for Quantum Science and Technology (MCQST), 80799 München, Germany}
\affiliation{Department of Mathematics, University of Houston, 77204 Houston, USA}
\author{Angelo Lucia}
\email{anglucia@ucm.es}
\author{David P\'{e}rez-Garc\'{i}a}
\email{dperezga@ucm.es}
\affiliation{Departamento de An\'{a}lisis Matemático y Matemática Aplicada, Universidad Complutense de Madrid, 28040 Madrid, Spain}
\affiliation{Instituto de Ciencias Matemáticas, 28049 Madrid, Spain}
\author{Cambyse Rouz\'{e}}
\email{cambyse.rouze@tum.de}
\affiliation{Munich Center for Quantum Science and Technology (MCQST), 80799 München, Germany}

\begin{abstract}
{We prove that spin chains weakly coupled to a large heat bath thermalize rapidly at any temperature for finite-range, translation-invariant commuting Hamiltonians, reaching equilibrium in a time which scales logarithmically with the system size.}
From a physical point of view, our result rigorously establishes the absence of dissipative phase transitions for Davies evolutions over translation-invariant spin chains. The result has also implications in the understanding of Symmetry Protected Topological phases for open quantum systems. 
\end{abstract}
\maketitle

\section{Introduction}

Understanding how thermal noise affects quantum systems is a major open problem in emerging quantum technologies. A key question there is {\it how long does it take for a system to thermalize?} i.e. to converge to its thermal Gibbs state. Or more specifically, which is the dependency of the thermalization time, also known as {\it mixing or decoherence time}, on the temperature and the system size? 

In particular, it is important to identify those scenarios in which the mixing time scales only logarithmically with the system size - such property is usually called  {\it rapid mixing}. From a negative point of view, in this regime quantum properties that hold in the ground state but not in the thermal state are suppressed too fast for them to be of any reasonable use. In the positive side, thermal states with such short mixing time can be constructed very efficiently with a quantum device that simulates the effect of the corresponding thermal bath. Let us note that constructing thermal Gibbs states is one of the main expected applications of a quantum computer, both as an important self-standing problem \cite{temme2011metropolis}, and also as a stepping stone in optimization problems, via simulated annealing type algorithms \cite{huang1987statisticalmechanics,somma2008annealing,brandao2017quantum,kieferova2017tomography,biamonte2017machinelearning}. On top of that, rapidly mixing systems have very desirable properties, such as stability with respect to extensive perturbations in the noise operator \cite{cubitt2015stability,Michalakis2015}.

Despite the importance of the question, very few mathematically rigorous results are known in this direction. The reason is the lack of mathematical techniques to tackle the problem. Indeed, the analogous results for classical systems already required sophisticated mathematical tools, in particular,  the notion of log-Sobolev constant for the noise infinitesimal generator, whereas estimates for the spectral gap of the generator are not enough to guarantee such rapid mixing. Starting with the pioneer work of Glauber for the particular case of classical 1D Ising model in 1963 \cite{glauber1963time}, Holley and Stroock \cite{holley1989uniform} (building upon a previous result of Holley \cite{holley1985rapid}) managed to prove the rapid mixing property for all 1D classical models at any temperature. This was done by showing that the log-Sobolev constant decreases at most logarithmically with the system size. Later, Zegarlinski \cite{zegarlinski1990log} improved their result showing that the log-Sobolev constant was indeed bounded.

In the quantum regime all results have focused on systems with commuting interactions. Note that this does not imply at all that the system is classical. Indeed, such systems include all types of non-chiral topological phases of matter. However most known results deal only with the spectral gap of the generator, what can only guarantee a mixing time that grows polynomially (and not logarithmically) with the system size \cite{kastoryano2016quantum}. For instance, Alicki et al. \cite{alicki2009thermalization} proved that the spectral gap has a uniform bound independent of the system size for the quantum ferromagnetic 1D Ising model and for Kitaev's Toric Code in 2D at all temperatures. This result was extended later for all abelian \cite{komar2016necessity} and non-abelian \cite{lucia2021thermalization} Kitaev's quantum double models in 2D, as well as for all 1D models with commuting interactions \cite{kastoryano2016quantum}.

In \cite{Gross1975}, a log-Sobolev inequality was also introduced in the quantum regime. In particular, a bounded (or logarithmically growing) associated constant is known to imply rapid mixing \cite{KastoryanoTemme-LogSobolevInequalities-2013,Pastawski2014}. Since then, several works have appeared trying to estimate such constant for different noise models in many body quantum systems with commuting interactions \cite{KastoryanoTemme-LogSobolevInequalities-2013,Pastawski2014,bardet2019modified,capel2020MLSI}. Despite considerable effort, the state of the art is that estimates have been obtained either for rather artificial noise operators \cite{capel2018quantum,Beigi2020} or only for sufficiently high temperature \cite{capel2020MLSI}.

In this Letter, we prove that \textbf{at any temperature, 1D quantum systems with commuting interactions are rapidly mixing}. That is, they thermalize in a time that scales only logarithmically with the system size. The approach to prove it is to bound the log-Sobolev constant of the associated Davies generator, which is the usual model for the action of a thermal bath in the weak coupling limit.

This result yields interesting consequences in the context of phase transitions. It is well known that the phases of a given system in thermal equilibrium can be classified according to its physical properties. Moreover, changes of the system allow for the transformation between phases, sometimes abruptly, which results in the appearance of a phase transition. This can also occur in systems which are away from their thermal equilibrium. In this case, due to dissipation, the environment drives the system to the aforementioned equilibrium, which is represented by a state and depends on the system and the environment parameters. As such parameters change, the properties of the system might also change suddenly, yielding a so-called \textit{dissipative phase transition} \cite{werner2005dissipativeising,capriotti2005dissipativephasetransitions,diehl2008phases,morrison2008phasetransitions,kessler2012dissipativephasetransitions}, sometimes also referred to as \textit{noise-driven quantum phase transitions} \cite{horstmann2013dissipativephasetransitions} or simply \textit{quantum phase transitions driven by dissipation} \cite{verstraete2009stateengineering}. Our result in particular implies that there are no dissipative phase transitions for Davies evolutions over translation-invariant spin chains.

The result also has implications in the context of Symmetry Protected Topological (SPT) phases \cite{wen2002quantum,gu2009tensor,chen2011classification}. There has been a quite intensive study of SPT in open quantum systems \cite{Diehl2011,rainis2012majorana,Viyuela2012,Bardyn2013,viyuela2015symmetry,Roberts2017,mcginley2019interacting,coser2019classification,mcginley2020fragility,degroot2021,Altland2021} and there is yet no consensus on what is the fate of SPT in the presence of temperature (see e.g. \cite{Viyuela2012, Roberts2017} for negative results and \cite{viyuela2015symmetry, viyuela2018observation} for positive ones). The 1D cluster state \cite{Briegel2001} - which plays a key role in the paradigm of measurement based quantum computation \cite{Raussendorf2001,Raussendorf2002} - has a commuting Hamiltonian and it is a non-trivial SPT phase under a $\mathbb{Z}_2\times \mathbb{Z}_2$ symmetry \cite{Son2011}. Hence, our result applies to it and gives the first example of a non-trivial interacting SPT phase with a provable decoherence time growing only logarithmically with the system size for thermal noise at every non-zero temperature, where in addition all relevant interactions in the problem can be asked to preserve the symmetry, at least in a weak sense. The case of a trivial Hamiltonian, which classically corresponds to the infinite-temperature case, was already obtained in \cite{coser2019classification}. The result has the extra benefit of being stable to extensive perturbations, a general property of quasi-local dissipative evolutions with logarithmic decoherence time \cite{cubitt2015stability}.  

Our result does not apply however in the presence of a strong symmetry \cite{buvca2012note,Albert2014}, a key condition identified in  \cite{degroot2021} to preserve SPT in open quantum systems, which emphasizes even more the totally different behavior between weak and strong symmetries in the context of SPT phases in non-zero temperature regimes.

The proof of our main result has two main steps. One step works in arbitrary dimension and gives a way to upgrade a bound on the spectral gap of the Davies generator to a bound of the log-Sobolev constant for commuting Hamiltonians. The proof requires among other things the theory of operator spaces, that has been already  proven very useful in answering different questions within quantum information theory \cite{palazuelos2016survey}. The other step is to show that 1D systems fulfill the hypothesis for such upgrade to hold. 

We expect the first step to be of independent interest, since it opens the possibility to upgrade to the log-Sobolev regime the recent result \cite{lucia2021thermalization} showing that the Davies generator of quantum double models in 2D have a bounded gap.

\section{Mixing times for Davies maps}

We now briefly recall the construction due to Davies \cite{Davies1974}, which under the assumption of a weak-coupling limit with a thermal bath at inverse temperature $\beta$, gives a description of the evolution of the system as a Markovian master equation. The joint Hamiltonian of the system and the environment can be decomposed as $H = H_S \otimes \identity_E + \identity_S \otimes H_E + \lambda H_I$, where $H_S$ is the Hamiltonian of the system, $H_E$ the one of the bath, and $H_I$ is the coupling term between the two of them, with the coupling constant $\lambda \ge 0$. We can decompose $H_I$ as $H_I = \sum_{\alpha} S^{\alpha} \otimes B^{\alpha}$, where $S^{\alpha}$, $B^{\alpha}$ are Hermitian. Renormalizing by the free evolution and sending $\lambda \to 0$ while keeping $\tau = \lambda^2 t$ constant, the reduced evolution of the system is given by $ \rho(\tau) = \exp( \tau \mathcal L)(\rho(0))$ \cite{Davies1974}. Here, $\cL$ is a Lindbladian whose Lindblad operators, which we denote by $S^\alpha(\omega)$, satisfy
$e^{itH_S} S^{\alpha} e^{-itH_S} = \sum_{\omega} S^\alpha(\omega) e^{-i \omega t}$, where the sum is over the Bohr frequencies $\omega$ of the system Hamiltonian $H_S$ (for more details, we refer the reader to our companion paper \cite{bardet2021MLSIDavies1D}). 

Under the assumption that there are no operators commuting with every $S^\alpha$ except the multiples of identity, one can show \cite{wolf2012quantum} that the Gibbs state of $H_S$ at inverse temperature $\beta$, namely $\sigma_\beta = {Z_\beta}^{-1} \exp(-\beta H_S)$
is the unique fixed point of the evolution generated by $\mathcal L$, and moreover:

\vspace{-10pt}
\begin{equation}\label{eq:convergence}
\forall\rho,~~    \exp(t \mathcal L)(\rho) \overset{t \rightarrow \infty}{\underset{\,}{\longrightarrow}} \sigma_\beta.
\end{equation}
\vspace{-10pt}

An important problem concerns the speed at which the convergence \eqref{eq:convergence} occurs. This is quantified by the \textit{mixing time} of the dynamics: for $\eps>0$,
\begin{align}
    t_{\operatorname{mix}}(\epsilon):= \inf\big\{t\ge 0|\, \|e^{t\cL}(\rho)-\sigma_\beta\|_1\le \eps\big\}\,,
\end{align}
where $\|X\|_1:=\tr|X|$ denotes the trace norm. One way of controlling the mixing time is via the analysis of the spectral gap of $\cL$. It is well-known that, whenever the gap can be lower bounded by a constant independent of system size $|\Lambda|$ \cite{KastoryanoTemme-LogSobolevInequalities-2013}, $t_{\operatorname{mix}}(\eps)=\mathcal{O}(\sqrt{n})$. This is the case for Davies generators over spin chains at any positive temperature \cite{kastoryano2016quantum}. Moreover, Glauber dynamics, which can be interpreted as the classical analogues of Davies generators, are known to thermalize logarithmically faster in 1D \cite{holley1989uniform,zegarlinski1990log} with $t_{\operatorname{mix}}(\eps)=\mathcal{O}(\operatorname{polylog}(n))$. This property of a local (quantum) Markovian evolution is known as \textit{rapid mixing}.

One way to prove rapid mixing is to consider the exponential decay of the relative entropy between the evolved state at time $t$ and the invariant state $\sigma_\beta$:
\begin{align}\label{eq:MLSI}
    D(e^{t\cL}(\rho)\|\sigma_\beta)\le e^{-4\alpha t}\,D(\rho\|\sigma_\beta)\,.
\end{align}
The constant $\alpha$ appearing in \eqref{eq:MLSI} is known as the \textit{modified logarithmic Sobolev constant} (MLSI constant) of the semigroup. By Pinsker's inequality together with the bound $D(\rho\|\sigma_\beta)=\mathcal{O}(\log(n))$, one can easily show that $\alpha=\Omega(\operatorname{polylog}(n)^{-1})$ implies the rapid mixing property. This is precisely what we achieve in this article.

\section{Main result}

We now state the main result of our paper, namely an exponential decay for the entropy in the form of Equation \eqref{eq:MLSI}. We consider a finite chain $\Lambda= \llbracket 1,n\rrbracket$ and the Davies generator  $\mathcal{L}_{\Lambda}$  of a quantum Markov semigroup with unique invariant state $\sigma \equiv \sigma_\Lambda^\beta := \frac{\operatorname{e}^{- \beta H_\Lambda}}{\tr[\operatorname{e}^{- \beta H_\Lambda}]}$, the Gibbs state of a finite-range, translation-invariant, commuting Hamiltonian at inverse temperature $\beta < \infty$.

\begin{theorem}\label{thm:main}
In the setting introduced above, there exists $\alpha_\Lambda=\Omega((\ln|\Lambda|)^{-1})$ such that, for all $\rho\in\cD(\cH_\Lambda)$ and all $t\ge 0$,
\begin{align}\label{eq:entropic_decay}
D(e^{t\cL_{\Lambda}}(\rho)\|\sigma)\le e^{-\alpha_\Lambda t}\,D(\rho\|\sigma)\,,
\end{align}
\end{theorem}

\begin{figure*}
\begin{center}
   \begin{tikzpicture}[scale=0.22]
   \begin{scope}
        \node [rectangle, aspect=2, draw, fill=lightblue!10,
     text width=13.2em, text centered, rounded corners, minimum height=3.5em,xshift=1.72cm,yshift=-0.05cm] {};
      \Block[17,coolblue!20!white,1][xshift=-5cm,yshift=-1cm];
            \node[black,xshift=-0.3cm,yshift=0.25cm] { \textbf{\textcolor{black}{a)}}};
      \node[black,xshift=1.5cm, yshift=-0.4cm] { \footnotesize{\textcolor{midblue}{$D(\rho \| \sigma)$}}};
   \end{scope}
      \begin{scope}[xshift=18.5cm]
     \draw [thick,black,decorate,decoration={brace,amplitude=16pt,mirror},xshift=-0.5pt,yshift=-0.6pt](3.1,10) -- (3.1,-8.8);
      \end{scope}
   \begin{scope}[xshift=25cm,yshift=6cm]
    \node [rectangle, aspect=2, draw, fill=lightblue!10,
     text width=13em, text centered, rounded corners, minimum height=7em,xshift=1.6cm,yshift=-0.3cm] {};
              \node[black,xshift=-0.35cm,yshift=0.6cm] { \textbf{\textcolor{black}{b)}}};
   \Block[5,midblue!60!white,1];
   \draw [thick,darkblue](-0.5,-0.5) -- (-0.5,0.5) ;
     \draw [thick,darkblue](4.5,-0.5) -- (4.5,0.5) ;
     \draw [thick,darkblue](-0.5,0.5) -- (4.5,0.5) ;
          \draw [thick,darkblue](-0.5,-0.5) -- (4.5,-0.5) ;
\draw [thick,midblue,decorate,decoration={brace,amplitude=6pt},xshift=-0.5pt,yshift=-0.6pt](-0.5,0.6) -- (4.5,0.6) node[black,midway,yshift=0.42cm,xshift=0.2cm] {  \footnotesize{\textcolor{midblue}{$D_{A_1}(\rho \| \sigma)$}}};
\node[black,xshift=1.5cm, yshift=-0.3cm] { \footnotesize \textbf{\textcolor{black}{+}}};
   \end{scope}
\begin{scope}[xshift=30cm,yshift=6cm]
\Block[3,coolblue!20!white,1];
  \draw [thick,darkblue](-0.5,-0.5) -- (-0.5,0.5) ;
\end{scope}
\begin{scope}[xshift=33cm,yshift=6cm]
\Block[5,midblue!60!white,1];
   \draw [thick,darkblue](-0.5,-0.5) -- (-0.5,0.5) ;
     \draw [thick,darkblue](4.5,-0.5) -- (4.5,0.5) ;
     \draw [thick,darkblue](-0.5,0.5) -- (4.5,0.5) ;
          \draw [thick,darkblue](-0.5,-0.5) -- (4.5,-0.5) ;
\draw [thick,midblue,decorate,decoration={brace,amplitude=6pt},xshift=-0.5pt,yshift=-0.6pt](-0.5,0.6) -- (4.5,0.6) node[black,midway,yshift=0.42cm,xshift=0.2cm] {  \footnotesize{\textcolor{midblue}{$D_{A_2}(\rho \| \sigma)$}}};
\end{scope}
\begin{scope}[xshift=38cm,yshift=6cm]
\Block[4,coolblue!20!white,1];
  \draw [thick,darkblue](-0.5,-0.5) -- (-0.5,0.5) ;
\end{scope}
 \begin{scope}[xshift=25cm,yshift=3cm]
   \Block[4,coolblue!20!white,1];
   \end{scope}
\begin{scope}[xshift=29cm,yshift=3cm]
\Block[5,midgreen!50!white,1];
   \draw [thick,darkgreen](-0.5,-0.5) -- (-0.5,0.5) ;
     \draw [thick,darkgreen](4.5,-0.5) -- (4.5,0.5) ;
     \draw [thick,darkgreen](-0.5,0.5) -- (4.5,0.5) ;
          \draw [thick,darkgreen](-0.5,-0.5) -- (4.5,-0.5) ;
\draw [thick,midgreen,decorate,decoration={brace,amplitude=6pt,mirror},xshift=-0.5pt,yshift=-0.6pt](-0.5,-0.6) -- (4.5,-0.6) node[black,midway,yshift=-0.4cm] {  \footnotesize{\textcolor{midgreen}{$D_{B_1}(\rho \| \sigma)$}}};
\end{scope}
\begin{scope}[xshift=34cm,yshift=3cm]
\Block[3,coolblue!20!white,1];
   \draw [thick,darkgreen](-0.5,-0.5) -- (-0.5,0.5) ;
\end{scope}
\begin{scope}[xshift=37cm,yshift=3cm]
\Block[5,midgreen!50!white,1];
   \draw [thick,darkgreen](-0.5,-0.5) -- (-0.5,0.5) ;
     \draw [thick,darkgreen](4.5,-0.5) -- (4.5,0.5) ;
     \draw [thick,darkgreen](-0.5,0.5) -- (4.5,0.5) ;
          \draw [thick,darkgreen](-0.5,-0.5) -- (4.5,-0.5) ;
\draw [thick,midgreen,decorate,decoration={brace,amplitude=6pt,mirror},xshift=-0.5pt,yshift=-0.6pt](-0.5,-0.6) -- (4.5,-0.6) node[black,midway,yshift=-0.4cm] {  \footnotesize{\textcolor{midgreen}{$D_{B_2}(\rho \| \sigma)$}}};
\end{scope}
\node at (17.2,1.3) {\large $\Lambda$};

\begin{scope}[xshift=37cm,yshift=3cm]
\node[black,xshift=-0.9cm, yshift=-1.15cm] { \textbf{\textcolor{black}{$\times$}}};
\end{scope}
  \begin{scope}[xshift=25cm,yshift=-5cm]
    \node [rectangle, aspect=2, draw, fill=lightblue!10,
     text width=13em, text centered, rounded corners, minimum height=3em,xshift=1.6cm,yshift=-0.3cm] {};
              \node[black,xshift=-0.35cm,yshift=-0.01cm] { \textbf{\textcolor{black}{c)}}};
   \Block[4,lightpurple!40!white,1];
\draw [thick,lightpurple,decorate,decoration={brace,amplitude=6pt,mirror},xshift=-0.5pt,yshift=-0.6pt](-0.5,-0.6) -- (3.5,-0.6) node[black,midway,yshift=-0.4cm] {  \footnotesize{\textcolor{lightpurple}{$\sigma_{B^c}$}}};
   \end{scope}
\begin{scope}[xshift=29cm,yshift=-5cm]
\Block[1,coolblue!20!white,1];
\draw[stealth-stealth](-0.8,-1.8) -- (0.8,-1.8);
\end{scope}
\begin{scope}[xshift=30cm,yshift=-5cm]
\Block[3,lightpink!40!white,1];
\draw [thick,lightpink,decorate,decoration={brace,amplitude=6pt,mirror},xshift=-0.5pt,yshift=-0.6pt](-0.5,-0.6) -- (2.5,-0.6) node[black,midway,yshift=-0.4cm] {  \footnotesize{\textcolor{lightpink}{$\sigma_{A^c}$}}};
\end{scope}
\begin{scope}[xshift=33cm,yshift=-5cm]
\Block[1,coolblue!20!white,1];
\draw[stealth-stealth](-0.8,-1.8) -- (0.8,-1.8);
\end{scope}
\begin{scope}[xshift=34cm,yshift=-5cm]
\Block[3,lightpurple!40!white,1];
\draw [thick,lightpurple,decorate,decoration={brace,amplitude=6pt,mirror},xshift=-0.5pt,yshift=-0.6pt](-0.5,-0.6) -- (2.5,-0.6) node[black,midway,yshift=-0.4cm] {  \footnotesize{\textcolor{lightpurple}{$\sigma_{B^c}$}}};
\end{scope}
\begin{scope}[xshift=37cm,yshift=-5cm]
\Block[1,coolblue!20!white,1];
\draw[stealth-stealth](-0.8,-1.8) -- (0.8,-1.8);
\end{scope}
\begin{scope}[xshift=38cm,yshift=-5cm]
\Block[4,lightpink!40!white,1];
\draw [thick,lightpink,decorate,decoration={brace,amplitude=6pt,mirror},xshift=-0.5pt,yshift=-0.6pt](-0.5,-0.6) -- (3.5,-0.6) node[black,midway,yshift=-0.4cm] {  \footnotesize{\textcolor{lightpink}{$\sigma_{A^c}$}}};
\end{scope}
      \begin{scope}[xshift=49cm,yshift=5cm]
       \draw [thick,black,decorate,decoration={brace,amplitude=10pt,mirror},xshift=-0.5pt,yshift=-0.6pt](-3.8,1.6) -- (-3.8,-3.8);
    \node [rectangle, aspect=2, draw, fill=lightblue!10,
     text width=13em, text centered, rounded corners, minimum height=3.5em,xshift=1.5cm,yshift=-0.22cm] {};
              \node[black,xshift=-0.43cm,yshift=0.12cm] { \textbf{\textcolor{black}{d)}}};
   \Block[17,red!40!white,1];
\node[black,xshift=1.5cm, yshift=-0.5cm] { \footnotesize \textbf{\textcolor{darkred}{$\underset{x \in \Lambda}{\sum} \, D \left( \rho \| E_x(\rho) \right)$}}};
\draw [thick,darkred](-0.5,-0.5) -- (-0.5,0.5) ;
\draw [thick,darkred](0.5,-0.5) -- (0.5,0.5) ;
\draw [thick,darkred](1.5,-0.5) -- (1.5,0.5) ;
\draw [thick,darkred](2.5,-0.5) -- (2.5,0.5) ;
\draw [thick,darkred](3.5,-0.5) -- (3.5,0.5) ;
\draw [thick,darkred](4.5,-0.5) -- (4.5,0.5) ;
\draw [thick,darkred](5.5,-0.5) -- (5.5,0.5) ;
\draw [thick,darkred](6.5,-0.5) -- (6.5,0.5) ;
\draw [thick,darkred](7.5,-0.5) -- (7.5,0.5) ;
\draw [thick,darkred](8.5,-0.5) -- (8.5,0.5) ;
\draw [thick,darkred](9.5,-0.5) -- (9.5,0.5) ;
\draw [thick,darkred](10.5,-0.5) -- (10.5,0.5) ;
\draw [thick,darkred](11.5,-0.5) -- (11.5,0.5) ;
\draw [thick,darkred](12.5,-0.5) -- (12.5,0.5) ;
\draw [thick,darkred](13.5,-0.5) -- (13.5,0.5) ;
\draw [thick,darkred](14.5,-0.5) -- (14.5,0.5) ;
\draw [thick,darkred](15.5,-0.5) -- (15.5,0.5) ;
\draw [thick,darkred](16.5,-0.5) -- (16.5,0.5) ;
\draw [thick,darkred](-0.5,0.5) -- (16.5,0.5) ;
\draw [thick,darkred](-0.5,-0.5) -- (16.5,-0.5) ;
   \end{scope}
\end{tikzpicture}
\end{center}
  \caption{
  Quasi-factorizations of the relative entropy used in the proof of Theorem \ref{thm:main}: First, we consider in \textbf{a)} the relative entropy between the evolved state and the equilibrium in $\Lambda$ and reduce it to some conditional relative entropies in smaller regions $\lbrace{A_i, B_i \rbrace}$ as in \textbf{b)}, in the spirit of the results of \cite{capel2018quantum,bardet2019modified,capel2020MLSI}, and a multiplicative error term in \textbf{c)}, depending on how correlations decay on $\sigma$ between $(\cup_i A_i)^c$ and $(\cup_i B_i)^c$, which can be interpreted as a mixing condition and is controlled using Araki's estimates \cite{araki1969gibbs} as in the recent \cite{bluhm2021exponential}. In  \textbf{d)}, we use operator spaces to lift the results of \cite{gao2021spectral} to non-tracial conditional expectations to further reduce the latter conditional relative entropies to another form of on-site conditional relative entropies.  }
  \label{fig:sketch}
\end{figure*}
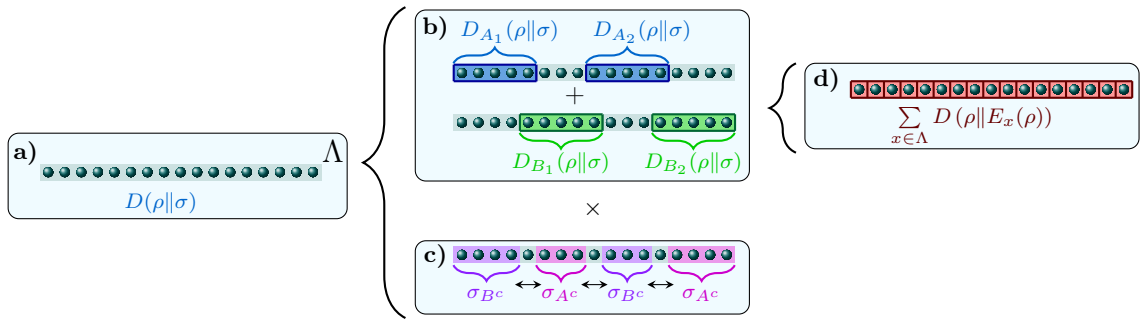

The proof of our result is schetched in Figure \ref{fig:sketch}. More details are provided in Appendix A. For a complete proof, we refer the reader to \cite{bardet2021MLSIDavies1D}.

\section{SPT phases}

In this section, we discuss how our result applies to spin chains that are preserved by certain symmetry of the system. We shall see that fast thermalization still holds even in the case of interactions with the thermal bath that preserve a symmetry of the system, thus negating the possibility of preserving information encoded in a thermal state. We begin with a brief recollection of properties of 1D SPT states before presenting our result.

\subsection{Definition}

Topological order is known to be robust against perturbations \cite{bravyi2010topological}, which endows those systems with a very desirable protection (e.g. for its use in quantum information processing tasks).  However, topological order is usually hard to engineer. In particular, it does not exist in 1D systems.

There is a somewhat related notion, where protection is guaranteed only if the perturbations keep a symmetry present in the system. Systems with that property are named Symmetry Protected Topologically (SPT) ordered \cite{gu2009tensor,wen2002quantum,chen2011classification}. SPT is a much less restricted property and there are several natural 1D systems which belong to a non-trivial SPT phase, among which the antiferromagnetic spin-$1$ Heisenberg model \cite{pollmann2010entanglement,haegeman2012order}.

In a series of papers \cite{chen2011classification,Schuch2011,Pollmann2012,fidkowski2011topological}, Matrix Product State (MPS) theory was used to classify all possible SPT phases in 1D via the second cohomology group $H^2(G, U(1))$, where $G$ is the symmetry group of the system. The idea is quite simple to explain. A MPS on a 1D system with local Hilbert space dimension $d$ is given by a sequence of $d$ matrices $A^i$, $i=1,\ldots,d$ of size $D\times D$, where $D$ is called the virtual or bond dimension. For a system of size $L$, the MPS defined by the $A^i$ is
\[|\psi_A\rangle = \sum_{i_1,\ldots, i_L} {\rm tr}\left(A^{i_1}A^{i_2} \cdots A^{i_L}\right)|i_1\cdots i_L\rangle \; .\]
That is, amplitudes are given by the trace of the product of matrices. A MPS is said to be \emph{injective} if there exists a finite $K$ such that $\{ A^{i_1}\cdots A^{i_K} \mid i_1,\dots,i_K=1\dots d\}$ spans the whole algebra of $D\times D$ matrices.

If a MPS is invariant under an on-site symmetry, i.e. $u_g^{\otimes L} |\psi_A\rangle= |\psi_A\rangle$ for all $L$ and all $g\in G$ -- $u_g$ is a given linear representation of $G$ -- then thanks to the Fundamental Theorem of MPS \cite{Sanz2009,Cirac2017,DelasCuevas2017}, there exists a {\it projective} representation $V_g$ of $G$ so that $\sum_{j} (u_g)_{ij} A^j = V_g A^i V_g^\dagger$.
That is, the action of the representation $u_g$ on the physical index of the $A^i$ (which is precisely the index $i$) corresponds to the adjoint action of the representation $V_g$ on the virtual indices (which are the row and column entries of the matrices $A^i$).

Non-trivial SPT phases correspond to the cases in which $V_g$ is {\it not} equivalent to a linear representation. Since the second cohomology group $H^2(G, U(1))$  enumerates all non-equivalent projective representations, it classifies non-trivial SPT phases associated to a symmetry group $G$.

It is by now well-known that MPS efficiently approximate ground states of locally interacting gapped Hamiltonians, what makes this classification, in principle restricted to MPS, valid in full generality, as first observed in numerical examples \cite{haegeman2012order} and then proven in full rigor by Ogata in \cite{ogata2021}.

\subsection{Example: the cluster state}

When the Hamiltonian has commuting terms, its ground state is an exact MPS and the argument given above works with full rigor. This is the case of the 1D cluster model, a Hamiltonian with 3-body terms between each particle and its nearest neighbors given by $Z\otimes X\otimes Z$, and where we are assuming periodic boundary conditions.

This Hamiltonian has a unique ground state which is precisely the 1D cluster state, the MPS given by matrices  $A^0=|0)(+|$, $A^1=|1)(-|$ \cite{cirac2020matrix} (we use curved-brackets to indicate that the associated spaces correspond to the virtual indices.)

If we block two sites, the corresponding matrices are $A^{00}=\frac{1}{\sqrt{2}} |0)(+|$,  $A^{01}=\frac{1}{\sqrt{2}} |0)(+|$, $A^{10}=\frac{1}{\sqrt{2}} |1)(+|$, and $A^{11}=-\frac{1}{\sqrt{2}} |1)(-|$.

It is straightforward that the 1D cluster Hamiltonian (on an even-size chain) commutes with the $\mathbb{Z}_2 \times \mathbb{Z}_2$ symmetry generated by $Z\otimes \mathbbm{1}$ and $\mathbbm{1}\otimes Z$. The MPS, being its unique ground state, must keep the same symmetry. Moreover, the $\mathbb{Z}_2 \times \mathbb{Z}_2$ symmetry  transfers to the virtual system: acting with $X\otimes \mathbbm{1}$ in the physical level corresponds to conjugation by $X$ in the virtual level, whereas acting with  $\mathbbm{1}\otimes  X$ in the physical level corresponds to conjugation by $Z$ in the virtual level.

Hence, the linear representation of $\mathbb{Z}_2 \times \mathbb{Z}_2$ in the physical level corresponds to the {\it projective} representation given by the Pauli group at the virtual level. The fact that such projective representation is {\it not} equivalent to a linear one (equivalently, it corresponds to the only non-trivial element of the second cohomology group $H^2(\mathbb{Z}_2 \times \mathbb{Z}_2, U(1))$), is precisely what gives the 1D cluster state its non-trivial SPT character.

\subsection{Davies generator for SPT phases}

In the case of an injective MPS invariant with respect to an on-site symmetry $u_g$ for some group $G$, one can also construct a local two-body frustration-free Hamiltonian $H_S$ which commutes with $u_g$, known as the \emph{parent Hamiltonian}
\[
    [H_S, u_g] = 0 \quad \forall g\in G.
\]
In this situation, we ask for the Davies thermalization process to also respect the symmetry $G$. We will do this by requiring that the Davies generator is \emph{covariant} with respect to the symmetry $u_g$: for every state $\rho$ and every $g \in G$, it holds that
\begin{equation} \label{eq:covariance-definition}
    \mathcal L(u_g^\dag \rho u_g) = u_g^\dag \mathcal L(\rho) u_g \quad \forall \rho, \forall g\in G.
\end{equation}
We remark that a sufficient condition for this to happen is for the jump operators $S^{\alpha}$'s to commute with $u_g$ up to a phase:
\begin{equation}\label{eq:covariance-condition}
    S^{\alpha} u_g = \omega^\alpha_g u_g S^{\alpha}, \quad \omega^\alpha_g \in U(1)\,.
\end{equation}
It is easy to construct many examples of covariant Davies generators. In fact, this is always possible when the parent Hamiltonian of the MPS is a sum of Pauli terms (tensor products of Pauli matrices), by choosing $S^\alpha$ to be the one-site Pauli operators. This covers the case of the 1D cluster state. 


We also remark that any such covariant generator $\mathcal L$ can be obtained as the weak-coupling limit of the interaction with a thermal bath that is \emph{weakly symmetric}, in the sense that there exists a representation $U_g$ of $G$ acting on the Hilbert space of the environment such that, for each $g\in G$ and all $\alpha$, $[H_E, U_g] = 0$ and $[S^\alpha\otimes B^\alpha, u_g \otimes U_g] = 0$. In fact, if this is not the case, one can extend the original environment by a conjugate copy of the system:
\[
    \tilde B^{\alpha} = B^\alpha \otimes \overline{S}^\alpha,\quad U_g = \identity \otimes \overline{u}_g,
\]
obtaining a weakly-symmetric thermal bath.


\subsection{Thermalization of 1D SPT phases}

Our main result applied to the 1D SPT phases state implies the following.

\begin{cor}
    For every inverse temperature $\beta <\infty$,
    1D SPT phases thermalize in time logarithmic in the system size, even when the thermal bath is chosen to be weakly symmetric.
\end{cor}

\section{Discussion}

In this work, we have shown that the Davies dynamics associated to any 1D spin chain translation-invariant commuting Hamiltonian at finite temperature satisfies a log-Sobolev inequality, and therefore the corresponding thermalization process converges logarithmically fast in terms of the system size (the \emph{rapid mixing} property). This also holds under the assumption that the evolution is weakly symmetric with respect to a given symmetry, for example in the case of SPT phases.

We expect our two-step proof strategy to be relevant in higher dimensions. We leave the study of log-Sobolev constants for Davies generators of 2D double models, whose gap was recently investigated in \cite{lucia2021thermalization}, to future work.

Finally, one could ask whether our result for SPT phases would apply to the setting where the thermal bath is chosen to be \emph{strongly symmetric}, in the sense that the representation $U_g$ acting on the Hilbert space of the environment is the trivial one. This is not the case, given that this condition prevents the thermal evolution to be ergodic, and in particular for it to have a unique invariant state. This can be seen by noticing that strong symmetry would imply that all $u_g$ are invariant, in the sense that $\tr\, [u_g\, \mathcal L(\rho)] = 0$ for any $\rho.$ A sufficient condition for this to happen is that $[S^\alpha,u_g]=0$ for each $\alpha$ and $g\in G$. In the presence of a full rank invariant state, this condition is also necessary \cite{Carbone2012}. When $u_g$ is not irreducible (which is the case for local on-site symmetries), this implies that $\mathcal L$ has multiple invariant states and therefore it is not ergodic. This issue was solved in \cite{Roberts2017} by restricting the initial state only to the subspace of $u_g$-symmetric state, and studying the thermalization of the \emph{symmetric Gibbs ensemble} (a non-full rank state). We leave open the question of whether our techniques could be adapted to cover this case.

\section*{Acknowledgements}
We are grateful for discussions with C. de Groot and A. Turzillo. IB is supported by French A.N.R. grant: ANR-20-CE47-0014-01 ``ESQuisses''. AC was partially supported by an MCQST Distinguished Postdoc and the Seed Funding Program of the MCQST (EXC-2111/Projekt-ID: 390814868). AL acknowledges support from the BBVA Fundation and the Spanish ``Ramón y Cajal'' Programme (RYC2019-026475-I / AEI / 10.13039/501100011033). This project has received funding from the European Research Council (ERC) under the European Union’s Horizon 2020 research and innovation programme (grant agreement No 648913).  DPG acknowledges support from MINECO (grant MTM2017-88385-P) and from Comunidad de Madrid (grant QUITEMAD-CM, ref. P2018/TCS-4342). CR acknowledges financial support from a Junior Researcher START Fellowship from the MCQST (EXC-2111/Projekt-ID: 390814868).

\bibliographystyle{unsrtnat}
\bibliography{references}

\appendix

\section{Techniques and proof ideas}\label{sec:proof_main_result}

In this appendix, we expose the main idea behind the proof of our main result, namely the exponential decay of the relative entropy in Equation \eqref{eq:entropic_decay}. For that, let us first rewrite such an equation in the following equivalent form:
\begin{equation}\label{eq:global_MLSI}
    2 \alpha D (\rho \| \sigma) \leq -\tr[\mathcal{L}_{\Lambda } (\rho) \left( \ln \rho - \ln \sigma \right)]
 \end{equation}
 Moreover, let us recall that we say that the Markovian evolution $( \operatorname{e}^{t \mathcal{L}_\Lambda})_{t \geq 0}$ satisfies a \textit{complete modified logarithmic Sobolev inequality} (CMLSI in short) if the $( \operatorname{e}^{t \mathcal{L}_\Lambda}\otimes \operatorname{id}_R)_{t \geq 0}$ has a positive MLSI constant $\alpha$ for any reference system $\mathcal{H}_R$.

The proof of Theorem \ref{thm:main} relies on a reduction argument from the MLSI constant of Equation \eqref{eq:global_MLSI} in $\Lambda$ to the complete MLSI constants in smaller sub-chains $A, B \subset \Lambda$. With this reduction we intend to overcome the issue that a MLSI constant in  $\Lambda$ could in principle scale with the system size. By relating it to complete MLSI constants in segments of fixed size, we expect that the original MLSI constant only depends on the size of such segments and, thus, is controllable.

This idea was originally introduced in \cite{cesi2001quasi, daipra2002classicalMLSI} in the context of classical spin systems, with the aim of simplifying the traditional strategy  for proving positivity of MLSI constants. Lately, a quantum analogue of such a procedure has been developed in  \cite{capel2018quantum, bardet2019modified, capel2019thesis, bardet2020approximate, capel2020MLSI, gao2021spectral}. The key element of this strategy are the so-called results of \textit{quasi-factorization} (or \textit{approximate tensorization}) of the relative entropy. The proof of Theorem \ref{thm:main} is also based on a result of this form.

The intuition behind this argument is the following: If both sides of inequality \eqref{eq:global_MLSI} were linear with respect to the lattice where they are considered, it would be a simple exercise to reduce an MLSI constant in $\Lambda= A \cup B$ to the two  MLSI constants in $ A$ and $B$. The RHS of Equation \eqref{eq:global_MLSI} actually satisfy such linearity. The generator that appears there is indeed local, in the sense that $\mathcal{L}_{\Lambda}={\sum}_{x \in \Lambda} \mathcal{L}_{x}$. Therefore, one can study the entropy production of $\rho$ in $A \subset \Lambda$ by restricting to the terms of the generator in $A$, i.e.
\begin{equation}\label{eq:linearity}
    \operatorname{EP}_A(\rho) := - \tr\left[\mathcal{L}_{A}(\rho) \left( \ln \rho - \ln \sigma \right)\right] \, ,
\end{equation}
for $\mathcal{L}_{A}(\rho) = {\sum}_{x \in A} \mathcal{L}_{x}(\rho)$. This yields the direct consequence that for two non-overlapping regions $A,B\subseteq \Lambda$,
\begin{align}\label{eq:additivity}
    \operatorname{EP}_{A\cup B}(\rho)=  \operatorname{EP}_{A}(\rho)+  \operatorname{EP}_{ B}(\rho)\,.
\end{align}
However, the LHS of Equation \eqref{eq:global_MLSI} is more subtle, as the relative entropy is clearly not linear with respect to the lattice where the states have support. Moreover, our aim at this stage is to reduce the area of action of the relative entropy, rather than that of the density matrices involved. This is the reason for introducing a couple of \textit{conditional relative entropies} in $A \subset \Lambda$ \cite{capel2018quantum,bardet2020approximate}:
for $\rho, \sigma \in \mathcal{D}(\mathcal{H})$:
\begin{align}
    D_A (\rho \| \sigma) & := D(\rho \| \sigma ) - D (\rho_{A^c} \| \sigma_{A^c}) \, , \\
      \mathbb{D}_A (\rho \| \sigma) & := D(\rho \| E_{A}(\rho)) \, ,
\end{align}
where $E_{A}$ is the conditional expectation such that $e^{t\cL_{A}}\to E_{A}$ as $t\to\infty$. These conditional relative entropies satisfy the following inequality \cite{bardet2020approximate}:
\begin{equation}\label{eq:ineq_conditional_relative_entropies}
    D_A (\rho \| \sigma) \leq \mathbb{D}_A (\rho \| \sigma)
\end{equation}
With these notions at hand, we provide an elaborate result of quasi-factorization of the relative entropy in terms of some conditional relative entropies in overlapping subsystems and a multiplicative error term which we can control. This is done in the two steps detailed below.

\subsection{Global-to-local reduction}

We first recall the following result of \textit{quasi-factorization} of the relative entropy \cite{capel2018quantum}: Consider a splitting of $\Lambda$ into two subchains $A$ and $B$ as in the following figure.
\begin{figure}[H]
\begin{center}
  \begin{tikzpicture}[scale=0.5]
\Block[6,midblue!60!white,1];
\begin{scope}
\draw [thick,midblue,decorate,decoration={brace,amplitude=10pt,mirror},xshift=-0.5pt,yshift=-0.6pt](-0.5,-0.6) -- (7.5,-0.6) node[black,midway,yshift=-0.7cm] { \textcolor{midblue}{$A$}};
\end{scope}
\begin{scope}[xshift=6cm]
\Block[2,coolgreen!50!white,1];
\end{scope}
\begin{scope}[xshift=8cm]
\Block[6,midgreen!50!white,1];
\draw [thick,midgreen,decorate,decoration={brace,amplitude=10pt},xshift=-0.5pt,yshift=-0.6pt](-2.5,0.6) -- (5.5,0.6) node[black,midway,yshift=0.7cm] { \textcolor{midgreen}{$B$}};
\end{scope}
\node at (14,1.2) {\Large $\Lambda$};
\end{tikzpicture}
\end{center}
  \caption{Possible splitting of a lattice $\Lambda$ into two subregions $A, B$ with non-overlapping complements and such that $\Lambda = A \cup B$.}
  \label{fig:0}
\end{figure}
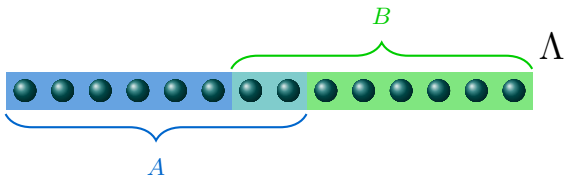
Given $\sigma  \in \cD(\cH_\Lambda)$, let us denote
\begin{align}\label{eq:definition_h_mixing_condition}
  &  h(\sigma_{A^cB^c})   \\
  & :=\sigma_{A^c}^{-1/2}\otimes \sigma_{B^c}^{-1/2}\sigma_{A^cB^c}\sigma_{A^c}^{-1/2}\otimes \sigma_{B^c}^{-1/2}-\Id_{A^cB^c} \, \nonumber.
\end{align}
Then, for all states $\rho, \sigma \in \cD(\cH_\Lambda)$ such that $\norm{ h(\sigma_{A^cB^c})}_\infty < 1/2$, the following holds:
\begin{align}\label{ATmixingcondrel}
  & D(\rho \| \sigma )  \\
  & \le \frac{1}{1-2\,\|h(\sigma_{A^cB^c})\|_\infty}\,\big[D_A(\rho \| \sigma )+D_B(\rho \| \sigma )\big]\,\nonumber .
\end{align}
This inequality between the relative entropy of two states in $\Lambda$ and the respective conditional relative entropies in $A$ and $B$ would allow us to reduce the MLSI constant in $\Lambda$ to the complete MLSI constants in $A$ and $B$. However, since $A$ and $B$ grow with $\Lambda$, this would still yield a bad scaling of the MLSI constant with the system size. To overcome this issue, we use a more involved geometry, by splitting our chain $\Lambda$ into small segments $\left\lbrace A_i\right\rbrace$ and  $\left\lbrace B_i\right\rbrace$ so that they present a pairwise overlap and all of them are of the same size, and consider $A:= \bigcup_i A_i$, $B:= \bigcup_i B_i$. This construction is as in the following picture.

\begin{figure}[H]
\begin{center}
  \begin{tikzpicture}[scale=0.42]
\Block[4,midblue!60!white,1];
\draw [thick,midblue,decorate,decoration={brace,amplitude=10pt,mirror},xshift=-0.5pt,yshift=-0.6pt](-0.5,-0.6) -- (4.5,-0.6) node[black,midway,yshift=-0.7cm] { \textcolor{midblue}{$A_1$}};
\begin{scope}[xshift=4cm]
\Block[1,coolgreen!45!white,1];
\end{scope}
\begin{scope}[xshift=5cm]
\Block[3,midgreen!50!white,1];
\draw [thick,midgreen,decorate,decoration={brace,amplitude=10pt},xshift=-0.5pt,yshift=-0.6pt](-1.5,0.6) -- (3.5,0.6) node[black,midway,yshift=+0.7cm] { \textcolor{midgreen}{$B_1$}};
\end{scope}
\begin{scope}[xshift=8cm]
\Block[1,coolgreen!45!white,1];
\end{scope}
\begin{scope}[xshift=9cm]
\Block[3,midblue!60!white,1];
\draw [thick,midblue,decorate,decoration={brace,amplitude=10pt,mirror},xshift=-0.5pt,yshift=-0.6pt](-1.5,-0.6) -- (3.5,-0.6) node[black,midway,yshift=-0.7cm] { \textcolor{midblue}{$A_2$}};
\end{scope}
\begin{scope}[xshift=12cm]
\Block[1,coolgreen!45!white,1];
\end{scope}
\begin{scope}[xshift=13cm]
\Block[4,midgreen!50!white,1];
\draw [thick,midgreen,decorate,decoration={brace,amplitude=10pt},xshift=-0.5pt,yshift=-0.6pt](-1.5,0.6) -- (3.5,0.6) node[black,midway,yshift=+0.7cm] { \textcolor{midgreen}{$B_2$}};
\end{scope}
\node at (17,1.2) {\Large $\Lambda$};
\end{tikzpicture}
\end{center}
  \caption{Splitting of $\Lambda$ into two regions $A$ and $B$, which are the union of small intervals  $\cup_{i=1}^m A_i$ and $\cup_{j=1}^m B_j$, such that $A_i \cap B_i \neq \emptyset \neq B_i \cap A_{i+1}$ for all $i = 1, \ldots m-1$ and subregions $A_i \cap A_j = \emptyset = B_i \cap B_j$ for all $i \neq j$.}
  \label{fig:1}
\end{figure}
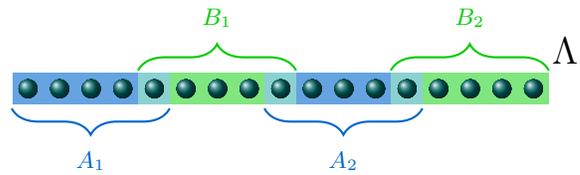
With this geometry at hand, we can further upper bound the conditional relative entropies in the RHS of Equation \eqref{ATmixingcondrel} by the conditional relative entropies on each $A_i$ and $B_i$ as a consequence of $\sigma$ being the Gibbs state of a local, commuting Hamiltonian, and, in particular, a quantum Markov chain \cite{bardet2019modified}:
\begin{align}\label{ATmixingcondrel2}
    &D(\rho \| \sigma )\le\\&  \frac{1}{1-2\,\|h(\sigma_{A^cB^c})\|_\infty}   \underset{i=1}{\overset{m}{\sum}} \big[D_{A_i}(\rho \| \sigma )+D_{B_i}(\rho \| \sigma )\big]\,\nonumber .
\end{align}
Next, we need to control the multiplicative error term in the RHS of Equation \eqref{ATmixingcondrel2}. Intuitively, the quantity $\|h(\sigma_{A^cB^c})\|_\infty$ should decrease with the size of the overlap between $A$ and $B$, as their complements get more separated. This can be proven to hold as a consequence of the recent \cite[Proposition 8.1]{bluhm2021exponential}, based on the so-called Araki's expansionals \cite{araki1969gibbs}. Indeed, for the geometry $\Lambda = A \cup B$ introduced above, we prove that $\|h(\sigma_{A^cB^c})\|_\infty$ is uniformly bounded by a constant away from $1/2$ as long as we take the number of segments $\left\lbrace A_i, B_i \right\rbrace$ to be $2m$, with $m = \Omega (\abs{\Lambda} / \ln (\abs{\Lambda}) )$, and the size of each segment $\abs{A_i}=\abs{B_i}=\mathcal{O}(\ln(\abs{\Lambda}) )$.

To conclude this step, we combine the previous inequalities with Equation \eqref{eq:ineq_conditional_relative_entropies} to obtain:
\begin{equation}\label{eq:AT}
\boxed{D(\rho \, \| \, \sigma)    \le \mathcal{C} \sum_{i=1}^{2m}D \big(\rho \,  \big\|E_{X_i}(\rho) \big)}
\end{equation}
for some universal constant $\mathcal{C} \ge 1$ and a covering $\{X_i\}_{i=1}^{2m}$ of the chain $\Lambda$ by intervals $X_i$ of size $|X_i|=O(\ln(\abs{\Lambda}))$.

\subsection{Quasi-local control of the constant}

In the next step, we reduce the conditional expectations on the regions $X_i$ in the relative entropies in the RHS of Equation \eqref{eq:AT} to single-site conditional expectations on each of the sites composing  each $X_i$. For that, we extend a previous result of quasi-factorization (also called approximate tensorization) for tracial conditional expectations from the recent \cite{gao2021spectral} to the non-tracial setting .

More specifically, for each $X_i$  we prove the following inequality
\begin{equation}\label{eq:corollaryAT}
    \boxed{D \big(\rho\|{E}_{X_i}(\rho)\big)\le \mathcal{K}_{X_i} \sum_{j\in X_i}\,D\big(\rho\|{E}_{j}(\rho)\big)}
    \end{equation}
    where $\mathcal{K}_{X_i}$ scales logarithmically with $\abs{\Lambda}$ whenever $|X_i| = \mathcal{O}(\ln \abs{\Lambda})$. Inspired from LaRacuente's work \cite{laracuente2019quasi}, we observe such constant $\mathcal{K}_{X_i}$ can be determined by the norm of the distance between the product of single-site conditional expectations ${E}_{j}$ and the conditional expectation ${E}_{X_i}$ for the region $X_i$. Indeed, using operator space theory we show that $\mathcal{K}_{X_i}$ is essentially the smallest integer $k$ such that
   $(\prod_{j\in X_i} E_j)^{k}$ is close enough to ${E}_{X_i}$ as maps between the amalgamated $L_p$ space introduced in \cite{junge2010mixed}. In finite dimensions, the norm of $(\prod_{j\in X_i} E_j)^{k}-{E}_{X_i}$ goes to $0$ as long as we can control the norm of $\prod_{j\in X_i} E_j-{E}_{X_i}$ on the $L_2$ spaces by a universal constant $\lambda<1$. This is obtained as a consequence of the non-closure of the spectral gap for 1D commuting Gibbs samplers proven in \cite{kastoryano2016quantum} and the detectability lemma \cite{aharonov2009detectability,aharonov2010quantum}. While the latter was already used in \cite{kastoryano2016quantum} to prove that a certain condition of exponential decay of correlations implies the non-closure of the gap, our main contribution here consists in leveraging this proof to get the approximate tensorization \eqref{eq:corollaryAT} with constant $\mathcal{K}_{X_i}\sim \Omega(\frac{|X_i|}{\ln \lambda})$. For this, we extend some of the results of \cite{gao2021spectral} to the Gibbs setting by developing the operator space structure of non-tracial amalgamated $L_p$ spaces. 
   
   The amalgamated $L_p$ norm has been previously used \cite{bardet2018hypercontractivity,gao2020fisher} in the study of MLSI constants of quantum Markov semigroups. They are generalizations of Pisier's vector noncommutative $L_p$ spaces \cite{pisier1998non}, which are closely related to sandwiched R\'enyi condition entropy. For more information on this, we refer the interested reader to our companion paper \cite{bardet2021MLSIDavies1D}. To the authors' best knowledge, this is the first time such techniques were used in the analysis of dissipative many-body quantum systems, and we strongly believe that they will find more applications elsewhere.

\subsection{Merging global and quasi-local analysis}

To conclude, we combine Equations \eqref{eq:AT} and \eqref{eq:corollaryAT} to obtain the following result of quasi-factorization of the relative entropy:
\begin{equation}\label{eq:general_AT}
    D(\rho \, \| \, \sigma)    \le \mathcal{K} \sum_{x \in \Lambda} D \big(\rho \,  \big\|E_{x}(\rho) \big)
\end{equation}
for $\mathcal{K}$ scaling logarithmically with $\abs{\Lambda}$. Now, we recall the positivity of the complete MLSI for the local generators $\mathcal{L}_x$ proven in \cite[Theorem 3.3]{gao2021spectral}: there exist a constant $\alpha_0 >0$ such that for all $x \in \Lambda$ and any $\rho \in \mathcal{D}(\mathcal{H}_\Lambda)$,
\begin{align}\label{eq:onesiteMLSI}
 \alpha_0   \,D(\rho\|E_{x}(\rho))\le \operatorname{EP}_{\{x\}}(\rho) \,.
\end{align}
Combining Equations \eqref{eq:general_AT}, \eqref{eq:onesiteMLSI} and the linearity of the entropy production \eqref{eq:linearity}, we directly get
\begin{align}
      D(\rho \, \| \, \sigma)    \le
        \mathcal{K}\alpha_0^{-1} \operatorname{EP}_\Lambda(\rho) \, ,
\end{align}
and the result follows.

\end{document}